\documentstyle[12pt]{article}
\newcommand{\bq}{\begin{equation}}
\newcommand{\eq}{\end{equation}}
\newcommand{\bqa}{\begin{eqnarray}}
\newcommand{\eqa}{\end{eqnarray}}
\newcommand{\ra}{\rightarrow}

\def\half{{1 \over 2}}
\def\s{\sigma}

\def\D{\Delta}
\def\a{\alpha}
\def\b{\beta}
\def\g{\gamma}

\def\l{\lambda}
\def\tp{2\pi i}
\def\ov{\over}

\def\rt{\sqrt{2}}
\def\ra{\rightarrow}

\def\2pi{1\over 2\pi i}
\def\q{q-q^{-1}}
\def\.{\mathaccent''005F}
\def\^{\mathaccent''007E}
\def\~{\tilde}
\def\newline{\hfil\break}

\def\ra{\rightarrow}

\def\sq2{\sqrt{2}}
\def\sqk2{\sqrt{2(k+2}}
\def\sqk{\sqrt{k}}
\def\sqs{\sqrt{2\over k}}

\def\be{\begin{equation}}
\def\ee{\end{equation}}
\def\br{\begin{array}}
\def\er{\end{array}}
\def\bea{\begin{eqnarray}}
\def\eea{\end{eqnarray}}
\newcommand{\uq}{U_q(\widehat{su}(2))}

\def\cF{{\cal{F}}}
\def\vp{\Omega}
\def\vP{\Phi}
\def\vt{\tilde{\Phi}}
\def\vo{\phi}
\def\cvo{\widehat{\phi}}
\begin{document}

\vbox{\vspace{15mm}}
\vspace{1.0truecm}
\begin{center}
{\LARGE \bf N-point matrix elements of  dynamical
 vertex operators of the higher spin XXZ model}\\[8mm]
{\large A.H. BOUGOURZI}\\
[3mm]{\it Centre de Recherches Math\'ematiques,
Universit\'e de Montr\'eal\\
C.P. 6128-A, Montr\'eal (Qu\'ebec) H3C 3J7, Canada\\
Email: bougourz@ere.umontreal.ca}\\[15mm]
\end{center}
\begin{abstract}
We extend the concept of conjugate vertex operators, first
introduced by Dotsenko in the case of the bosonization of the
$su(2)$ conformal field theory, to
the bosonization of the dynamical vertex operators
(type II in the classification
of the Kyoto school)
of the higher spin XXZ model. We show that the
introduction of the conjugate vertex operators leads to
simpler expressions for the N-point matrix
elements of the dynamical vertex operators, that is, without  redundant
Jackson integrals that arise from the insertion of screening charges.
In particular, the two-point
matrix element  can be represented without any  integral.

\end{abstract}
\newpage
\section{Introduction \label{secv1}}

It is  now well established  through the work of the Kyoto school
that the XXZ quantum spin chain with local spins equal to $k/2$
has the quantum affine algebra $\uq$ with level $k$
as a non-Abelian symmetry; $q$ being a deformation
(anisotropy) parameter
\cite{book,Daval92,qaffine}.  The dynamical symmetry of this model is generated
by the dynamical vertex operators (referred to as type II vertex operators in
Refs. \cite{book,Daval92,qaffine}) which create non-degenerate
 eigenstates of the XXZ Hamiltonian by a successive action on a
given eigenvector, for example the vacuum. A dynamical
vertex operator has two
main properties: it intertwines the $\uq$ modules and  carries spin 1/2
\cite{qaffine}.

An important mathematical and physical quantity in this model
is the N-point matrix elements of the above vertex
operators. There are three known ways to compute these matrix elements
exactly, at least in principle. The first one is by solving
 the q-KZ equation. Indeed, it has been
shown in Ref. \cite{FrRe92} that these vertex operators
satisfy a difference equation, which is the q-analogue
of the usual KZ equation in conformal field theory.
The second one consists of deriving a normal
ordering of the modes of the vertex operators that
is compatible with the Zamolodchikov-Fateev algebra they
satisfy. With this normal ordering one can then
in principle compute any matrix element.
The third one consists in  realizing the vertex operators in terms of
bosonic modes satisfying Heisenberg algebras. Since
the normal ordering of the latter modes is very simple, one can use it to
compute  matrix elements.

In practice however, the first two methods are hardly useful
beyond the two-point matrix elements because of highly technical
complications, whereas the third one, though more useful and
systematic raises also the
following problem: a single vertex operators might
have several independent realizations in terms of bosonic modes.
Equivalently, a given $\uq$ representation might be identified with  several
Fock spaces  with different bosonic charges (eigenvalues with respect to the
center of the Heisenberg algebra).
Therefore, it is not trivial which combinations
of these realizations of
the vertex operators and Fock spaces are going to lead to the right matrix
elements.
This problem has been addressed in the context of the bosonization of
conformal field theory
through two different approaches \cite{Dot90}, which we extend
to  the case of $\uq$ algebra.
The first one consists of singling out only one  particular realization,
say the simplest,
for the vertex operators  and then attaching an appropriate
number of  screening charges to these vertex operators
so that the resulting operator
is a map connecting to the two Fock spaces between which we aim
to compute the matrix elements. This amounts to fixing
the same bosonic charge (picture fixing)
for both Fock spaces realizing
a $\uq$ representation and its dual.
The main disadvantage of this method is that the final
expressions for the matrix elements  involve in general redundant
integrations coming from the screening
charges. The second method, which is due to Dotsenko (in the classical case
i.e., $q=1$) \cite{Dot90},
consists of deriving all  possible independent bosonizations of the
vertex operators.
Then one should single out two among them such that
the two-point matrix element can be computed as the expectation value
of the product of the two (i.e., each one of them is inserted once)
and without the insertion of
screening charges.
This is equivalent to fixing two different bosonic charges: one for
  the Fock space realizing
a $\uq$ representation and the other for the Fock space realizing
its dual.
This is the reason why
Dotsenko refers to
one of them as a vertex operator and the other one as its
conjugate vertex operator, i.e., they create respectively
a representation and its dual space. From this bosonic
realization of the two-point matrix element one reads off the
conservation law of the bosonic charges. This conservation
law must be obviously satisfied in the N-point matrix elements
otherwise one must again insert the minimum number of
screening charges in the matrix elements to make  it so.
 This second method has the
main advantage of avoiding unnecessary redundant integrations in the
integral representations of the matrix elements obtained through
bosonization.

So far only the first method has been applied in the
computation of matrix elements of the dynamical
vertex operators of the spin $k/2$ XXZ model \cite{Kon94}.
In this paper, we apply the second method to the computation
of these N-point matrix elements.
In section 2, we briefly review the $\uq$ algebra and its
bosonic realization. In section 3, we recall the definition
of the dynamical vertex operators as intertwiners of
$\uq$ modules. The bosonization of these vertex operators
are derived by solving the intertwining relations they
satisfy with the $\uq$ algebra, which is already in a
bosonized form. We show that here also there are  two independent
solutions, just as in conformal field theory \cite{Dot90}.
One of them has already been derived in Ref.
\cite{Kon94}, whereas the second one is new.
We refer to the
first solution as the vertex operator and to the second as
the  conjugate vertex operator. In section 4,  we
show how the N-point vacuum-to-vacuum matrix element
can be computed without
a redundant integration if both the vertex operators and
conjugate vertex operators are used simultaneously.
As an explicit example, we compute
the two-point matrix element in a non-integral form and show that
it satisfies the q-KZ equation. Finally section 5 is devoted
to our conclusions.

\section{The $U_q(\widehat{su(2)})$ algebra and its bosonization\label{secbos}}
The associative  unital $\uq$ algebra is generated by the elements
$\{ E^{\pm}_n~(n\in {\bf Z}),H_m ~(m \in {\bf Z}\backslash\{ 0\}),K^{\pm 1},
\g^{\pm 1/2}\}$, with the following defining relations  \cite{Dri85}:
\be\br{rcl}
&&KK^{-1}=K^{-1}K=1,\\
&&\g^{1\over 2}\g^{-{1\over 2}}=\g^{-{1\over 2}}\g^{1\over 2}=1,\\
&&{[K^{\pm 1}, H_m]}=  0,\\
&&  K  E^\pm_n K^{-1} =q^{\pm 2} E^\pm_n,\\
&&{[H_n,H_m]} = {[2n]\over 2n} {{\g^{n}-\g^{-n}} \ov {q-q^{-1}}}
\delta_{n+m,0},\\
&& {[H_n,E^{\pm}_m]}=
\pm\sqrt{2}{\g^{\mp |n|/2}[2n]\over 2n}
E^\pm_{n+m}, \\
&&{[E^+_n,E^-_m]} = {\g^{(n-m)/2}\psi_{n+m}-\g^{(m-n)/2}
\varphi_{n+m}\over q-q^{-1}},\\
&&E^\pm_{n+1}E^\pm_m-q^{\pm 2}E^\pm_mE^\pm_{n+1}=
q^{\pm 2}E^\pm
_nE^\pm_{m+1}-E^\pm_{m+1}E^\pm_n,
\label{cwb}
\er\ee
where $\g^{1/2}$ is in the centre of the algebra
and acts as $q^{k/ 2}$ on level $k$
highest weight representations of $\uq$, and
$\psi_n$ and $\varphi_n$ are
the modes of the fields
$\psi(z)$ and $\varphi(z)$ defined by
\be\br{rcl}
\psi(z)&=&\sum\limits_{n\geq 0}\psi_nz^{-n}=q^{\sqrt{2}H_0}
 \exp\{\sqrt{2}(\q)\sum\limits_{n>0}H_nz^{-n}\},\\
\varphi(z)&=&\sum\limits_{n\leq 0}\varphi_nz^{-n}=q^{-\sqrt{2}H_0}
\exp\{-\sqrt{2}(\q)\sum\limits_{n<0}H_nz^{-n}\}.
\label{algebra}\er\ee
As usual, $[x]$ is defined by $[x]=(q^x-q^{-x})/(q-q^{-1})$
and $q$ is the deformation parameter.
The above algebra is a Hopf algebra with the following comultiplication:
\newpage
\be\br{rcl}
\!\!\!\!\!\!& &\Delta(E^+_n)=E^+_n\otimes\gamma^{n}+
\gamma^{2n}K\otimes
E^+_n+ \sum_{i=0}^{n-1}\gamma^{(n+3i)/2}\psi_{n-i}\otimes \gamma^{n-i}
 E^+_i\: {\rm mod}   \: {N_-}\otimes {N_+^2},\\
 \!\!\!\!\!\!& &\Delta(E^+_{-m})=E^+_{-m}\!\otimes\!\gamma^{-m}\!+\!
K^{-1}\!\otimes\! E^+_{-m}+
\sum_{i=0}^{m-1}\gamma^{(m-i)/2}\varphi_{-m+i}\otimes
\gamma^{i-m}
 E^+_{-i}
\: {\rm mod  }\:
 N_-\otimes N_+^2,\\
\!\!\!\!\!\!& &\Delta(E^-_{-n})=E^-_{-n}\!\otimes\!\gamma^{-2n}K^{-1}\!+\!
  \gamma^{-n}\!\otimes \!E^-_{-n}\!+\!\!\sum_{i=0}^{n-1}\!\gamma^{i-n}
  E^-_i\!\otimes\!\gamma^{-(n+3i)/2}\varphi_{i-n}\: {\rm mod  }\:
 N_-^2\!\otimes\! N_+,\\
\!\!\!\!\!\!& &\Delta(E^-_m)=\gamma^{m}\otimes E^-_m+E^-_m\otimes
K +
\sum_{i=1}^{m-1}\gamma^{m-1}E^-_m\otimes \gamma^{(i-m)/2}
\psi_{m-i}\:{\rm mod  }\: N_-^2\otimes N_+,\\
\!\!\!\!\!\!& &\Delta(H_m)=H_m\otimes\gamma^{m/2}+
\gamma^{3m/2}\otimes H_m\: {\rm mod  }\: N_-\otimes N_+,\quad\\
\!\!\!\!\!\!&
&\Delta(H_{-m})=H_{-m}\otimes\gamma^{-3m/2}+\gamma^{-m/2}
\otimes  H_{-m}\: {\rm mod  }\: N_-\otimes N_+,\quad\\
\!\!\!\!\!\!& &\Delta(K^{\pm 1})=K^{\pm 1}\otimes
 K^{\pm 1},\quad\\
\!\!\!\!\!\!& &\Delta(\gamma^{\pm \half})=\gamma^{\pm\half}\otimes
\gamma^{\pm \half},
\label{comult}\er\ee
where $m>0$, $n\geq 0$, and $N_\pm$ and $N_\pm^2$ are
left ${\bf Q}[\gamma^\pm,  \psi_m, \varphi_{-n}; \: m, n\in {\bf
Z_{\geq 0}}]$-modules
generated
 by $\{E^\pm_m; \:m\in {\bf Z}\}$
and $\{E^\pm_m E^\pm_n; \:m, n\in {\bf Z}\}$  respectively
\cite{ChPr91}.
This comultiplication will be useful in deriving the intertwining properties of
the
vertex operators.

Let us now briefly review the bosonization of $\uq$
for arbitrary level $k$.
We need three deformed Heisenberg algebras generated by
the elements $\{a^j_n,  j=1,2,3;\quad n\in\bf Z\}$, to which we adjoin the
elements $a^j,j=1,2,3$, and  with the following
defining relations  \cite{Bou93}:
\be\br{rcl}
{[a^j_n,a^\ell_m]}&=&(-1)^{j-1}nI_j(n)\delta^{j,\ell}
\delta_{n+m,0},\\
 {[a^j,a^\ell_0]}&=&(-1)^{j-1}i\delta^{j,\ell},
\qquad\qquad\qquad  j,\ell=1,2,3,
\label{Heisen}
\er\ee
and where
\be\br{rcl}
I_1(n)&=&{[2n][nk]\over 2kn^2},\\
I_2(n)&=&{[nk][n(2+k)]\over n^2k(2+k)}q^{|n|k},\\
I_3(n)&=&{[2n]^2\over 4n^2}.
\er\ee
The bosonization of $\uq$ is
give by  \cite{Bou93}
\be\br{rcl}
\!\!\!\!\psi(z)&= &\exp\left\{
i\sqs\left(\chi^{1,+}(zq^{k/2})-\chi^{1,-}(zq^{-k/2})\right)
\right\}\\
\!\!\!\!\varphi(z)&= &\exp\left\{
i\sqs\left(\chi^{1,+}(zq^{-k/2})-\chi^{1,-}(zq^{k/2})\right)
\right\}\\
\!\!\!\!E^{+}(z)
&=& {1 \over {z(q-q^{-1})}} (E_+^{+}(z)-E_-^{+}(z)),\\
E^+_{\pm}(z)&=&\exp\{i\sqs\chi^{1,+}(z)+i\sqrt{{2+k\over k}}
\chi^2(z)-i\chi^3(zq^{\mp 1}) \},\\
\!\!\!\!E^{-}(z)
&=& {1 \over {z(q-q^{-1})}} (E_+^{-}(z)-E_-^{-}(z)),\\
E^-_{\pm}(z)&=&\exp\{- i\sqs\chi^{1,-}(z)
-i\sqrt{{2+k\over k}}
\chi^2(zq^{\pm k})+i\chi^3(zq^{\pm (1+k)})\},
\label{fif}
\er\ee
where $z$ is a complex variable and
\be\br{rcl}
E^{\pm}(z)&=&\sum_{n\in\bf Z}E^{\pm}_nz^{-n-1},\\
\chi^{1,\pm}(z)&=&\vp^1(k,1;1|0,\mp k/2,0|+|z),\\
\chi^{2}(z)&=&\vp^2(k,1;1|0,-k/2,0|+|z),\\
\chi^{3}(z)&=&\vp^3(2,1;1|0,0,k+2|+|z).
\er\ee
Here, we have used the notation
\be\br{lll}
\vp^j(L_1,L_2,\dots L_r;M_1,M_2,\dots,M_s|\s,\a,\b|\pm|z)&=
a^j-ia_0^j\ln{(\pm zq^\s})\\
&+i {{L_1L_2\dots L_r}\over M_1M_2\dots M_s}\sum\limits_{n\in{\bf Z}
\backslash\{0\}}
{{[M_1 n][M_2 n]\dots [M_s n]}\over {[L_1 n][L_2 n]\dots[L_r n]}}
a_n^j q^{\a |n|+\b n} z^{-n}.
\er\ee
where $L_1,L_2,\dots L_r$, $M_1,M_2,\dots,M_s$, $\sigma$,
$\alpha$ and $\beta$ are parameters
related to the q-deformation.

\section{Bosonization of the dynamical vertex operators}
Here we consider the bosonization of the dynamical
vertex operators and introduce the concept of
their conjugate vertex operators. These vertex operators
are referred to as type II vertex operators in Refs.
\cite{qaffine}. They map the $\uq$ modules
 in the following way:
\bq \vP^{j~j_2}_{j_1}(z):V(\Lambda_{j_1}) \ra V^j(z) \otimes V(\Lambda_{j_2}),
\label{origvo}\eq
where $V(\Lambda_j)$ are level $k$ highest weight
$\uq$-modules, and $\{\Lambda_j=(k-2j)\l_0+2j\l_1,~ j=0,\dots,k/2\}$
and $\{\l_0,\l_1\}$ denote the sets of $\uq$ dominant highest
weights and fundamental weights respectively. $V^j(z)$ is the
$k=0$ `evaluation representation' of $\uq$.
It is isomorphic to $V^j\otimes {\bf C}[z,z^{-1}]$, where $V^j$ is the
$2j+1$ dimensional
representation with the basis $\{v_m^j,~ -j\leq m\leq j\}$, and
 is equipped with the following $\uq$-module
structure:
\be\br{rcl}
\gamma^{\pm 1/2}v^j_m\otimes z^\ell&=&v^j_m\otimes z^\ell,\\
q^{\rt H_0} v_m^j\otimes z^\ell&=&q^{-2m}v^j_m\otimes z^\ell,\\
E^+_nv^j_m\otimes z^\ell&=&q^{2n(1-m)}[j+m]v^j_{m-1}\otimes z^{\ell+n},\\
E^-_nv^j_m\otimes z^\ell&=&q^{-2nm}[j-m]v^j_{m+1}\otimes z^{\ell+n},\\
H_nv^j_m\otimes z^\ell&=&{1\over{\rt n}}\{[2nj]-q^{n(j-m+1)}(q^n+q^{-n})
[n(j+m)]\}v^j_m\otimes z^{\ell+n},
\er\ee
with $v^j_m$ being identically zero if
$|m|> j$.

Let us introduce the rescaled vertex operators
$\vt^{j~j_2}_{j_1}(z)$ as
\be \vP^{j~j_2}_{j_1}(z)=z^{(\D_{j_2}-\D_{j_1})}
\vt^{j~j_2}_{j_1}(z),\ee
where $\D_j= j(j+1)/(k+2)$. The latter vertex operators
are defined to obey the following intertwining relations \cite{FrRe92,qaffine}:
\bq \vt^{j~j_2}_{j_1}(z) \circ x = \Delta(x)\circ
\vt^{j~j_2}_{j_1}(z) ~~~~~\forall~ x\in \uq,
\label{inter} \ee
where $\Delta$ is the comultiplication given in (\ref{comult}).
We define the components $\vo^j_m(z)$ of these vertex operators as
\be \vt^{j~j_2}_{j_1}(z) = g^{j~j_2}_{j_1}(z)
\sum\limits_{m=-j}^{j}  \vo^j_m(z)\otimes v^{j}_m ,\ee
where the normalization function $g^{j~j_2}_{~~j_1}(z)$ is to be
determined so that
\be
\vt^{j~j_2}_{j_1}(z)|\Lambda_{j_1}>=|\Lambda_{j_2}>+\dots,
\label{norm1}
\ee
with $|\Lambda_{j_1}>$ and $|\Lambda_{j_2}>$ being the highest weight states
of $V(\Lambda_{j_1})$ and $V(\Lambda_{j_2})$ respectively.

Using  relation (\ref{inter}), the
comultiplication (\ref{comult}), and the fact that
$N_+v^j_{-j}=N_-v^j_j=0$, $N_{\pm}v^j_m\in
{\bf C}[z,z^{-1}]
v^j_{m\mp 1}$, we get the following commutation relations:
\bea
{[E^-_n,\vo^j_{-j}(z)]}&=&0,\qquad n\in\bf Z,\label{a1}\\
{[H_n, \vo^j_{-j}(z)]}&=&
- q^{k(n-|n|/2)} { {[2jn]}\ov {\rt n} }z^n
\vo^j_{-j}(z),\qquad n\in{\bf Z}\backslash \{0\},\label{a2}\\
K\vo^j_{-j}(z)K^{-1}&=&q^{-2j}\vo^j_{-j}(z),\label{a3}\\
\vo^j_{m+1}(z)&=&{1\over [j+m+1]}[\vo^j_m(z),E^{+}_0]_{q^{-2m}},
\label{phijm}\eea
where the quantum commutator $[A,B]_{x}$ is defined by
\be\br{rcl}
[A,B]_{x}=AB-xBA.
\er\ee

As in the case of type I vertex operators \cite{BoWe941},
the system of equations (\ref{a1})-(\ref{a3})
has two independent solutions for $\vo^j_{-j}(z)$
in terms of the bosonic Heisenberg elements (\ref{Heisen}).
We
denote them respectively by $\vo^j_{-j}(z)$ and
$\cvo^j_{-j}(z)$.  They are given by
\be\br{rcl}
\vo^j_{-j}(z)&=&\exp\{-ij\sqrt{2\over k}\xi^1(z)
-{2ij(k+1)\over
\sqrt{k(k+2)}}\xi^2(z)+2ij\xi^3(z)\},\\
\cvo^j_{-j}(z)&=&\exp\{-ij\sqrt{2\over k}\xi^1(z)+
{i(k-2j)\over \sqrt{k(k+2)}}\widehat\xi^2(z)\},
\er\ee
where
\be\br{rlc}
&\xi^1(z)=\vp^1(2,k;2j|k,-k/2,-k|-|z),\\
&\xi^2(z)=\vp^2(k+2,k,1;2j,k+1|k,-k/2,-k|-|z),\\
&{\widehat\xi}^2(z)=\vp^2(k+2,k;k-2j|k,-k/2,-k|-|z),\\
&{\widehat\xi}^3(z)=\vp^3(2,1;2j|k,0,2|-|z).
\er\ee
The components $\vo^j_m(z)$ and $\cvo^j_m(z)$ are derived from
$\vo^j_{-j}(z)$ and $\cvo^j_{-j}(z)$ through (\ref{phijm}). Henceforth,
we will refer
to $\vo^j_m(z)$ and $\cvo^j_m(z)$ as  vertex operators  and
conjugate vertex operators  respectively.

Let us now define the  Fock spaces on which the
above operators are acting. A left Fock module $F(n_1,n_2,n_3)$
and a right Fock module $\bar F(n_1,n_2,n_3)$
labelled by the integers $n_1$, $n_2$ and $n_3$ are defined by
\be\br{rcl}
F(n_1,n_2,n_3)&=&F_-|n_1,n_2,n_3\rangle,\\
F_+|n_1,n_2,n_3\rangle&=&0\\
\bar F(n_1,n_2,n_3)&=&\langle n_1,n_2,n_3|F_+,\\
\langle n_1,n_2,n_3|F_-&=&0,
\er\ee
where $F_\mp$ are respectively generated by
$\{a_{\mp n}^1,a_{\mp n}^2,a_{\mp n}^3,~ n>0\}$. The states
$|n_1,n_2,n_3\rangle$ and $\langle n_1,n_2,n_3|$ are
defined by
\be\br{rcl}
|n_1,n_2,n_3\rangle=\exp\{ {n_1  \over \sqrt{2k}}ia^1+{n_2 \over
\sqrt{k(k+2)}}ia^2+n_3 ia^3\}|0\rangle,\\
\langle n_1,n_2,n_3|=\langle
0|\exp\{ - {n_1  \over \sqrt{2k}}ia^1-{n_2 \over
\sqrt{k(k+2)}}ia^2-n_3 ia^3\},
\er\ee
where the vacua states $|0\rangle=|0,0,0\rangle$ and $\langle 0|=\langle
0,0,0|$
are such that
\be\br{rcl}
a_n^i|0\rangle&=&0,\quad n\geq 0, \quad i=1,2,3,\\
\langle 0|a_n^i&=&0,\quad n\leq 0, \quad i=1,2,3.
\er\ee
{}From the  intertwining relations
 (\ref{a1})-(\ref{a3}) it is clear that the
following states are $\uq$ lowest
weight states:
\be\br{rcl}
|-2j,-2j(k+1),2j\rangle&=&:\phi^j_{-j}(0):|0\rangle=
\exp\{ -{2j  \over \sqrt{2k}}ia^1-{2j(k+1) \over
\sqrt{k(k+2)}}ia^2+2j ia^3\}|0\rangle,\\
|-2j,k-2j,0\rangle&=&:\widehat\phi^j_{-j}(0):|0\rangle=
\exp\{ -{2j  \over \sqrt{2k}}ia^1+{k-2j \over
\sqrt{k(k+2)}}ia^2\}|0\rangle.
\er\ee
As usual the symbol :: denotes the bosonic normal ordering, which is
defined so that the modes $\{a^i_n;n\geq 0,i=1,2,3\}$ are always
placed to the right of the  modes $\{a^i_n;n<0,i=1,2,3\}$.
Moreover, one can easily check that
$(E^+_0)^{2j}$ maps the state $|-2j,-2j(k+1),2j\rangle$
to  $|2j, 2j, 0\rangle$, which is a $\uq$
highest weight state.  The states $|2j, 2j, 0\rangle$ and
$|-2j, k-2j,0\rangle$
define respectively the
Fock modules $F(2j,2j,0)$ and
$F(-2j,k-2j,0)$.

The currents $E^\pm(z)$ and the vertex operators  $\vo^j_{m}(z)$ and
$\cvo^j_{m}(z)$  map
a Fock module $F(n_1,n_2,n_3)$ as:
\bea
\!\!\!E^{\pm}(z):& F(n_1,n_2,n_3)\ra F(n_1\pm 2,n_2\pm(k+2),n_3\mp 1),
\label{eact}\\
\!\!\!\vo^j_m(z):& F(n_1,n_2,n_3)\rightarrow F(n_1+2m,n_2+m(k+2)-jk,n_3+j-m),
\label{act2}\\
\!\!\!\cvo^j_m(z):& F(n_1,n_2,n_3)\rightarrow F(n_1+2m,n_2+k(j+1)+m(k+2),
n_3-j-m).
\label{action}\eea
One can show that the action of the currents is well defined (single valued)
on the Fock modules $F(n_1,n_2,n_3)$  provided
that the following
condition is satisfied:
\be
n_1-n_2\in k{\bf Z}.
\ee
Relation (\ref{eact}) implies that the representations on
which the currents are acting are the
complete Fock spaces
\bea
\cF (j)&=&\bigoplus_{r\in{\bf Z}}
F(2j+2r,2j+r(k+2),-r),\\
\widehat\cF (j)&=&\bigoplus_{r\in{\bf Z}}
F(-2j+2r,k-2j+r(k+2),-r). \label{FS}
\eea
Note however that the irreducible $\uq$ highest (lowest) weight representation
is  only a subspace embedded
in the  Fock space $\cF (j)$ ($\widehat\cF (j)$) and can be projected
out through a BRST analysis \cite{Kon93}.
We claim that the dual representation of the highest weight Fock space
$\cF (j)$ is isomorphic as a $\uq$ module to the lowest weight Fock space
$\widehat\cF (j)$. Moreover, we make the normalization
\bea
\langle 2j,2j(k+1)+k,-2j|2j,2j,0\rangle=1.
\label{nor}\eea
As far as we know, there is no algebraic proof of this claim
even in the classical case,
i.e., the $su(2)$ conformal field theory \cite{Dot90}
(see however Ref. \cite{Fel89} for a proof
in the case of the Virasoro algebra).
We will nevertheless check its validity through
the two-point matrix element.

The next ingredient we need is the notion of a
screening operator denoted by
$S(z)$. This is an operator that commutes with the currents
$E^\pm(z)$, $\psi(z)$ and $\phi(z)$ up to total quantum
derivatives. According to  Refs.
\cite{Mat922, BoWe942},  there are three such
operators in the case of $\uq$. However, for our purposes
here we  need only one of them. It is given by
\be\br{rcl}
S(z)~= ~{_1{\cal D}}_z(\exp\{-i{\eta}^3(z)\})
\exp\{i\sqrt{k\over k+2}{\eta}^2(z)\}
\er\ee
with
\be\br{rcl}
\eta^2(z)=\Omega^2(k+2,1;1|-k-2,-k/2,0|+|z),\\
\eta^3(z)=\Omega^3(2,1;1|-k-2,0,k+2|+|z),
\er\ee
and satisfies the following commutation relations:
\be\br{rcl}
{[\varphi(z),S(w)]}&=&0,\\
{[\psi(z),S(w)]}&=&0,\\
{[E^+(z),S(w)]}&=&0,\\
{[E^-(z),S(w)]}&=&-_{k+2}{\cal D}_w({h(w)\over z-w}),
\label{ecoms}\er\ee
for some operator $h(w)$.
The quantum derivative
$_k{\cal D}_zf(z)$ of a function $f(z)$ is as usual defined by:
\be\br{rcl}
_k{\cal D}_zf(z)={f(zq^k)-f(zq^{-k})\over z(q-q^{-1})}.
\er\ee
Relations (\ref{ecoms})
imply that the screening charge $Q$, which is constructed
as a Jackson
integral of the
 screening current, i.e.,
\be\br{rcl}
Q=\int_0^{s\infty}d_pz S(z) \label{Q}
\er\ee
with $p=q^{2(k+2)}$, commutes with $\uq$.
The Jackson integral of a function $f(z)$ is
defined by \cite{FrRe92}:
\be
\int_0^{s\infty}d_pz f(z)=s(1-p)\sum\limits_{n\in {\bf Z}}f(sp^n)p^n. \ee
The action of the screening charges (\ref{Q}) on
a Fock module $F(n_1,n_2,n_3)$ is given
by
\be Q:F(n_1,n_2,n_3) \ra F(n_1,n_2+k,n_3-1)\label{qact}.\ee
Because of the above properties of $Q$ and the relations
(\ref{act2})-(\ref{action}), one
can construct the following two screened vertex operators,
which map $\cF(j_1)$  to
$\cF(j_2)$ and $\widehat\cF(j_2)$ respectively,
\bea
\vt^{j~j_2}_{j_1}(z) &=& g^{j~j_2}_{j_1}(z)
\sum\limits_{m=-j}^{j} Q^r \vo^j_m(z)\otimes  v^{j}_m,\\
\widehat{\vt}^{j~j_2}_{j_1}(z) &=& \~g^{j~j_2}_{j_1}(z)
\sum\limits_{m=-j}^{j} Q^s \cvo^j_m(z)\otimes  v^{j}_m,
\eea
where $r=j+j_1-j_2$, $s=j_1+j_2-j$, and
$g^{j~j_2}_{j_1}(z)$ and $\~g^{j~j_2}_{j_1}(z)$ are
 normalization functions that can be
determined from  (\ref{norm1}).
Note that we can also construct two more screened vertex operators
that map $\widehat\cF(j_1)$ to $\cF(j_2)$ and $\widehat\cF(j_2)$, but since we
do
not need them in the sequel we will not consider them here.

\section{N-point matrix element}

The physically interesting situation
is when the vertex operators carry spin $j=1/2$, in which
case we denote them by
$\vP^{j_2}_{j_1}(z) \equiv\vP^{{1\over 2}~j_2}_{j_1}(z)$.
They map $\cF(j_1)$  to
$\cF(j_1 \pm 1/2)$ and $\widehat\cF(j_1 \pm 1/2)$.
The N-point matrix element we are interested in is
the vacuum-to-vacuum  expectation
value
$\langle 0|\vP^{j_n}_{j_{n-1}}(z_n)\vP^{j_{n-1}}_{j_{n-2}}(z_{n-1})
\cdots\vP^{j_2}_{j_{1}}(z_2)
\vP^{j_1}_{j_{0}}(z_0)|0\rangle$. Here $|0\rangle$ is identified with
the highest weight vector of $V(\Lambda_0)$.
This matrix
element is represented by  the following sum of matrix
elements in the Fock spaces:
\be\br{rcl}
&&\langle 0|\vP^{j_n}_{j_{n-1}}(z_n)\vP^{j_{n-1}}_{j_{n-2}}(z_{n-1})
\cdots\vP^{j_2}_{j_{1}}(z_2)
\vP^{j_1}_{j_{0}}(z_1)|0\rangle=
\sum\limits_{\{m_1,\cdots,m_n\}} \left\{\left(\prod\limits_{i=1}^{n}
z_i^{\D_{J_i}-\D_{J_{i-1}}} g^{j_i}_{j_{i-1}}(z_i)\right)\right.\\
&&\left.\times\langle 0|Q^{s_{n}}\cvo^{j}_{m_n}(z_n)
Q^{r_{n-1}}\vo^{j}_{m_{n-1}}(z_{n-1})
\cdots
Q^{r_{2}}\vo^{j}_{m_{2}}(z_{2})
Q^{r_{1}}\vo^{j}_{m_1}(z_1)
|0\rangle v_{m_n}^{j_n}\otimes \cdots \otimes
v_{m_1}^{j_1}\right\},\label{npoint}
\er\ee
where
\be\br{rcl}
j&=&1/2,\\
j_0&=&j_n=0,\\
j_1&=&j_{n-1}=1/2\\
j_i&=& j_{i-1}\pm 1/2,\qquad i=1,\dots,n\\
r_1&=&s_n=0\\
r_i&=&j_{i-1}-j_i+1/2,\qquad i=1,\dots,n-1.
\er\ee
The normalization constants $g^{j_i}_{j_{i-1}}(z)\equiv
g^{{1\over 2}~j_2}_{j_1}(z)$
for $i=1,\dots,n-1$ have been calculated
in Ref. \cite{Kon94}  and
$g^{j_n}_{j_{n-1}}(z)=\~g^{0}_{1/2}(z)\equiv \~g^{{1\over 2}~j_2}_{j_1}(z)$
is computed from the
requirement that
\bea
\~g^{0}_{1/2}(z)\cvo^{1/2}_{-1/2}(z)|\Lambda_1\rangle=
|\Lambda_0\rangle+\dots.
\eea
Here the state $|\Lambda_0\rangle=|0,k,0\rangle$ is the highest weight of
$V(\Lambda_0)$ which is embedded in $\widehat\cF (0)$. We find
\bea
\~g^{0}_{1/2}(z)=(-zq^k)^{3\over 2(k+2)}.
\label{g0}
\eea
Note that the consistency between the relations
(\ref{npoint}) and (\ref{nor}) imposes the following
two constraints:
\be\br{rcl}
\sum_{i=1}^n m_i&=&0,\\
L&\equiv& \sum_{i=1}^{n-1}r_i+s_n=j(n-2)={n-2\over 2},
\label{kar}\er\ee
where $L$ denotes the total number of screening charges.
This means that the only non-vanishing vacuum-to-vacuum
matrix elements are those of an even number of vertex operators. Moreover,
should one  use a vertex operator instead of a conjugate vertex operator in
(\ref{npoint}), and the same bosonic vacuum for both a $\uq$ representation
and its dual, i.e.,
$\langle 0|0\rangle =1$, then
the above constraints would have become
\be\br{rcl}
\sum_{i=1}^n m_i&=&0,\\
L&\equiv& \sum_{i=1}^{n}r_i=jn={n\over 2}.
\er\ee
Therefore by using the conjugate vertex operator we have removed one redundant
integral.
Since all the operators inside this expectation value
are realized explicitly in terms of bosonic modes, one can easily
normal order them by computing
all the operator product expansions (OPE's). The
c-function that we get from  these OPE's is precisely
the  expectation value
we are interested in
(the vacuum-to-vacuum matrix element of the
normal ordered exponential operators is equal to 1). One should note
however, that the resulting integral expression of this
expectation value is not always well defined because
the integrands are not necessarily convergent due to the
singularities that arise in the OPE's and clearly a kind
of regularization to remove any divergence is needed.
This issue has been addressed partially in Ref. \cite{Kon94}
but further analysis is required.

A particular case where the above problem does not arise
because simply the insertion of screening charges is not required
(indeed, if $n=2$ then relation (\ref{kar}) implies that $L=0$),
and where the computation can be made more
explicit, is provided by the two-point matrix element
$\langle 0|\vP^{0}_{1/2}(z)\vP^{1/2}_{0}(w)|0\rangle$. This
expectation value is represented as a vacuum expectation value
over the Fock spaces by
\be\br{rcl}
\langle 0|\vP^{0}_{1/2}(z)\vP^{1/2}_{0}(w)|0\rangle&=&z^{\D_{0}-\D_{1/2}}
w^{\D_{1/2}-\D_{0}}
\~g^{0}_{1/2}(z)g^{1/2}_{0}(w)\times\\
&&
\left(\langle 0|\cvo_{+}(z)\vo_{-}(w)|0\rangle v_{+}\otimes v_{-}
+\langle 0|\cvo_{-}(z)\vo_{+}(w)|0\rangle v_{-}\otimes v_{+}\right),
\label{2point}
\er\ee
with
\be\br{rcl}
\vo_{\pm}(z)&\equiv &\vo_{\pm 1/2}^{1/2}(z),\\
\cvo_{\pm}(z)&\equiv &\cvo_{\pm 1/2}^{1/2}(z),\\
v_{\pm}&\equiv& v^{1/2}_{\pm 1/2},\\
\D_0&=&0,\\
\D_{1/2}&=&{3\over 4(k+2)}.
\er\ee
$\~g^{0}_{1/2}(z)$ is given by (\ref{g0}) whereas
$g^{1/2}_{0}(z)$ can be obtained from the normalization
\bea
g^{1/2}_{0}(z)\vo_{+}(z)|\Lambda_0\rangle=
|\Lambda_1\rangle+\dots.
\eea
Using the explicit form of $\vo_+(z)$, which is given
below we find
\bea
g^{1/2}_{0}(z)=-q^{-1}.
\eea
The explicit bosonic constructions of
 the vertex operators
$\vo_{\pm}(z)$ and $\cvo_{\pm}(z)$
that we obtain by solving the intertwining
conditions (\ref{inter}) are given by
\be\br{rcl}
\vo_{-}(z) &=& \exp\left(-{i \over {\sqrt{2k}}}
\xi^1(z)-{i(k+1) \over {\sqrt{k(k+2)}}} \xi^2(z)
+i\xi^3(z)\right), \\
\vo_{+}(z)&=&[\vo_-(z), E^+_0]_q,\\
&=&q^{k+2}\oint\limits_{|w|<|zq^{1+k}|}
{dw \over \tp} {z:\vo_{-}(z)E_+^{+}(w):\ov w(w-zq^{1+k})}
-q^{k}\oint\limits_{|w|>|zq^{k-1}|}
{dw \over \tp} {z:\vo_{-}(z)E_-^{+}(w):\ov w(w-zq^{k-1})}
,\\
\cvo_{-}(z)&=&\exp\left(-{i \over {\sqrt{2k}}}
\xi^1(z)+{i(k-1) \over {\sqrt{k(k+2)}}} \widehat\xi^2(z)\right), \\
\cvo_{+}(z)&=&[\cvo_-(z), E^+_0]_q,\\
&=&q\oint\limits_{|w|<|zq^{1+k}|}
{dw \over \tp} {:\cvo_{-}(z)E^{+}(w):\ov w-zq^{1+k}}
-q\oint\limits_{|w|>|zq^{k-1}|}
{dw \over \tp} {:\vo_{-}(z)E^{+}(w):\ov w-zq^{k-1}}.
\label{cvominus}
\er\ee
In deriving relations (\ref{cvominus}) we have used
the following OPE's:
\be\br{rcl}
E^+_+(z)\vo_-(w)&=&q:E^+_+(z)\vo_-(w):,\\
\vo_-(z)E^+_+(w)&=&{z-wq^{1-k}
\over z-wq^{-k-1}}:\vo_-(z)E^+_+(w):,\qquad |wq^{-k-1}|<|z|,\\
E^+_+(z)\cvo_-(w)&=&{1
\over z-wq^{k-1}}:E^+_+(z)\cvo_-(w):,\qquad |wq^{k-1}|<|z|\\
\cvo_-(z)E^+_+(w)&=&-{q^{-k}
\over z-wq^{-k-1}}:E^+_+(z)\cvo_-(w):,\qquad |wq^{-k-1}|<|z|,\\
E^+_-(z)\vo_-(w)&=&q^{-1}{z-wq^{k+1}
\over z-wq^{k-1}}:E^+_-(z)\vo_-(w):,\qquad |wq^{k-1}|<|z|,\\
\vo_-(z)E^+_-(w)&=&:\vo_-(z)E^+_-(w):,\\
E^+_-(z)\cvo_-(w)&=&{1
\over z-wq^{k-1}}:E^+_-(z)\cvo_-(w):,\qquad |wq^{k-1}|<|z|,\\
\cvo_-(z)E^+_-(w)&=&-{q^{-k}
\over w-zq^{-k-1}}:E^+_-(z)\vo_-(w):,\qquad |wq^{-k-1}|<|z|.
\label{opep}
\er\ee
Note that when $k=1$ the conjugate vertex operator
$\cvo_-(z)$ satisfies exactly the same OPE's with
the currents $E^{\pm}(z)$
as those of the  vertex operator which is bosonized
through the Frenkel-Jing realization of $\uq$ \cite{Nak93,book}.
In this sense, unlike  $\vo_{\pm}(z)$,
$\cvo_{\pm}(z)$ are the natural generalizations of the vertex operators
found in the case  $k=1$ in Ref. \cite{Nak93}.
In order to compute the above two-point matrix element we need to use,
besides the OPE's (\ref{opep}), the following
OPE:
\be
\cvo_-(z)\vo_-(w)=-q^{k}(-zq^k)^{-{3\over 2(k+2)}}
(z-wq^{-2})
\prod\limits_{n=0}^{\infty} { {(p^{n} z/w)_{\infty}
(p^{n} q^4 z/w)_{\infty}} \ov {(p^{n}q^{-2} z/w)_{\infty}
(p^{n} q^6 z/w)_{\infty}} }:\cvo_-(z)\vo_-(w): \label{fzw},\ee
where $p=q^{2(k+2)}$ and
\be (a)_{\infty}=\prod\limits_{n=0}^{\infty}(1-aq^{4n}).\ee
Using all the OPE's given by (\ref{opep}) and (\ref{fzw}),
and carrying out the integrals we find
that the above two-point matrix element takes the
following simple form:
\be
\langle 0|\vP^{0}_{1/2}(z)\vP^{1/2}_{{0}}(w)|0\rangle=
f({w / z})(v_{-}\otimes v_{+} - q^{-1} v_{+}\otimes v_{-}),
\label{twopm}
\ee
where
\be
f(z)=z^{3\ov {4(k+2)}}\prod\limits_{n=0}^{\infty} { {(p^{n} z)_{\infty}
(p^{n} q^4 z)_{\infty}} \ov {(p^{n}q^{-2} z)_{\infty}
(p^{n} q^6 z)_{\infty}} }.\ee
It can be easily checked that the two-point matrix element
$\langle 0|\vP^{0}_{1/2}(z)\vP^{1/2}_{{0}}(w)|0\rangle$ as given in
(\ref{twopm})
 satisfies  the following q-KZ equations \cite{FrRe92,qaffine,Daval92}:
\be\br{rcl}
\langle 0|\vP^{0}_{1/2}(z)\vP^{1/2}_{{0}}(pw)|0\rangle&=&
(1\otimes K^{-1})R^+(w/z)\langle 0|\vP^{0}_{1/2}(z)\vP^{1/2}_{{0}}(w)|0
\rangle,\\
\langle 0|\vP^{0}_{1/2}(pz)\vP^{1/2}_{{0}}(pw)|0\rangle &=&
(K^{-1}\otimes K^{-1})\langle 0|\vP^{0}_{1/2}(z)\vP^{1/2}_{{0}}(w)|0\rangle,
\er\ee
where the $R$ matrix $R^+(z)$ is given by
\be
R^+(z)=\rho(z)\bar R(z),
\ee
with
\be\br{rcl}
\rho(z)&=&q^{-{1\over 2}}{(q^2 z)^2_\infty\over (z)_\infty (q^4z)_\infty},\\
\bar R(z)v_{\pm}\otimes v_{\pm}&=&v_{\pm}\otimes v_{\pm},\\
\bar R(z)v_{+}\otimes v_{-}&=&{(1-z)q\over 1-q^2 z}v_{+}\otimes v_{-}
+{(1-q^2)z\over 1-q^2 z}v_{-}\otimes v_{+},\\
\bar R(z)v_{-}\otimes v_{+}&=&{1-q^2\over 1-q^2 z}v_{+}\otimes v_{-}
+{(1-z)q\over 1-q^2 z}v_{-}\otimes v_{+}.
\er\ee
Moreover, when we set $k=1$, this two-point matrix element reduces to
\be \langle 0|\vP^{0}_{1/2}(z)\vP^{1/2}_{{0}}(w)|0\rangle=
(w/ z)^{1\ov 4}{ {(w/z)_{\infty}} \ov {(q^{-2}w/z)_{\infty}} }
(v_{-}\otimes v_{+} - q^{-1} v_{+}\otimes v_{-}),\ee
which is just the result found in Ref. \cite{Daval92} up to a similarity
transformation of the $R^+(z)$ matrix by the transposition matrix $P$, which
is defined by
\be\br{rcl}
P(v_{\pm}\otimes v_{\pm})&=&v_{\pm}\otimes v_{\pm},\\
P(v_{\pm}\otimes v_{\mp})&=&v_{\mp}\otimes v_{\pm}.
\er\ee

\section{Conclusions}

In this paper we derive two independent bosonizations of
each dynamical
vertex operator of the higher spin XXZ model.
Both of them satisfy the same
intertwining
relations. We refer to one of them as the vertex operator and the
other as its conjugate vertex operator. This terminology is due to our
 claim that when the former acts on the left and
the latter acts on the right
bosonic vacua,
they create respectively a $\uq$ left representation
and its dual.  We confirm the  validity of this
claim through
the two-point matrix elements. As a result, the N-point
matrix elements of dynamical vertex operators do not involve redundant
Jackson  integrals coming from the insertion of
screening charges. It would be interesting to find an algebraic proof to the
above
claim.
Finally , it is natural to
expect that the
use of conjugate vertex operators would also simplify the computation of
other important physical quantities like form factors
(traces of products of type I and type II vertex operators)
and correlation
functions (traces of type I vertex operators) of local operators
of the higher spin XXZ model. This is presently under investigation.

\section*{Acknowledgements}
We thank  M.A. El Gradechi, L. Vinet and R.A. Weston for
their useful comments on this
paper. We would like to thank T. Miwa for providing us with the unpublished
notes in Ref. \cite{Nak93}. We are also grateful to NSERC for
providing us with a postdoctoral fellowship.

\end{document}